\documentclass[review]{elsarticle}

\usepackage{amssymb}
\usepackage{amsmath, mathrsfs, amssymb,amsfonts,amsthm,graphicx, epsf, dcolumn}
\usepackage[hyperfootnotes=true]{hyperref}
\usepackage{subfigure}
\usepackage{color}
\usepackage{slashed}
\usepackage{setspace}
\usepackage{cancel}
\usepackage{wasysym}
\usepackage{hyperref}
\usepackage{float}
\usepackage{array}
\pdfoutput=1

\journal{Physics of the Dark Universe}

\begin{document}

\begin{frontmatter}



\title{On the viability of the evolution of the universe with Geometric Inflation}


\author{Luisa G. Jaime}
\address{Departamento de F\'isica, Facultad de Ciencias. Universidad Nacional Aut\'onoma de M\'exico, A.P. 70-543, CDMX, 04510, Ciudad Universitaria,  M\'exico}

\begin{abstract}

We perform a general analysis of the cosmological viability of Geometric Inflation.  We show that the evolution of the universe, from inflation to the present day, can be seen from the addition of an infinite tower of curvature invariants into the Hilbert-Einstein action. The main epochs of the Universe can be reproduced: Inflation, Big Bang Nucleosynthesis, and Late-time acceleration driven by the cosmological constant. The slow-roll condition is a robust prediction of the theory.  Inflation possesses a graceful exit with enough number of $e-$folds between the limit imposed by the Planck density and the exit of the exponential expansion to solve the horizon problem and the absence of topological defects.  We also provide some scenarios where the energy scale of the theory can be calibrated.
\end{abstract}




\end{frontmatter}


\section{Introduction}
\label{secc:Introduction}
Geometric modifications of gravity can take as its motivation a generalization of one of the hypotheses assumed in General Relativity (GR), such as metricity, torsion-free, and as in the case of $f(R)$, generalizations of the dependence on the Ricci scalar $R$ into the Hilbert-Einstein action (HEA) (see \cite{Heisenberg_2019} for a complete review of geometric modifications of gravity in cosmology). $f(R)$ gravity has been proposed in two cosmological scenarios, for inflation with the Starobinsky's addition of $R^2$ \cite{1980PhLB...91...99S}, and for late-time acceleration where some authors have introduced some functions of $R$ \cite{2007JETPL..86..157S, 2007PhRvD..76f4004H}. Currently, no proposal can unify both, inflation and late-time acceleration.

In the proposal of Geometric Inflation, we expand the generalization of the dependence of the Ricci scalar into the HEA by adding an infinite tower of curvature scalars. In \cite{Arciniega:2018fxj} we add all the cubic curvature invariants choosing Lagrangian densities in such a way that the theory is Einstenial-like and the field equations for a homogeneous and isotropic universe are of second order. This modification can produce an accelerated expansion at the beginning of the universe. Nevertheless, the expansion scales as a power law not exponentially. In \cite{Arciniega:2018tnn, Arciniega:2019oxa} we show that the addition of an infinite tower of curvature invariants can produce an inflationary period with a graceful exit \citep{Arciniega:2020pcy}. We have coined this modification as Geometric Inflation.

It is a well-known fact that inflationary models, based on the inflaton hypothesis \cite{2009arXiv0907.5424B}, require fine-tuning to fulfill the flatness and horizon conditions. In the present work, we show that, within the frame of Geometric Inflation, the slow-roll condition, $\epsilon =\dot{H}/H^2\ll 1$, is reached in the pass by imposing initial conditions at the present time. The horizon problem can be solved with enough number of $e-$folds during inflation. The graceful exit becomes natural and the exponential expansion is driven by radiation. We also explore the evolution of the universe under this theory during the time that is expected to occur Big Bang Nucleosynthesis. Finally, we revise the late-time acceleration driven by the cosmological constant and the consequences on the sound horizon scale.

\section{The Theory}
\label{TheTheory}
Geometric inflation (GI) theory consists of a modification into the HEA by adding an infinite tower of higher-order invariants in such a way that the Lagrangian density terms, order by order, will reproduce the Einsteinian characteristics (see below properties  $i)$ and $iii)$). The first analysis for the cosmology of this kind of modification was presented, for the cubic case, in \cite{Arciniega:2018fxj} where the early acceleration with graceful exit can be noticed. Nevertheless, the inflationary epoch presents a power-law expansion. In \cite{Arciniega:2018tnn} we introduce an infinite tower of curvature invariants and the expansion, during the inflationary period, becomes exponential with graceful exit. 

The modified HEA in Geometric inflation is 
\begin{equation}\label{theo}
S_{GI}=\int \frac{d^4x \sqrt{|g|}}{16\pi G}\left\{-2\Lambda+R+\sum_{n=3}^{\infty}\lambda_n L^{2n-2}\mathcal{R}_{(n)}\right\}\, ,
\end{equation}
where the term $\mathcal{R}_{(n)}$ represents Lagrangian densities involving contractions of the Riemann tensor until $n$-order. $\lambda_n$ are dimensionless couplings and $L$ represents an energy scale. The general properties of the theory that have been found are
\begin{itemize}
\item[i) ] Its vacumm spectrum solely consist of a graviton \cite{Bueno2016b}.  
\item[ii) ] On maximally symmetric space-times, the field equations are of second order \cite{Bueno2016b}.
\item[iii) ] Black holes solutions are non-hairy and Schwarszchild-like with $g_{tt}=g_{rr}^{-1}$ \cite{Bueno2017}.
\item[iv) ] Friedmann-Lema\^{i}tre-Robertson-Walker (FLRW) solutions have a well-possed initial value problem \citep{Arciniega:2018tnn}.
\item[v) ] Cosmological FLRW solutions produce an inflationary epoch gracefuly conected with the late-Universe \citep{Arciniega:2018tnn}.
\end{itemize}
In the present work, we are interested in points $iv)$ and $v)$, and we add some other relevant features of the theory to this list.
\section{Generalized Friedmann Equations}
\label{Friedmann eq}
\vskip1mm
For a FLRW Universe given by the metric
\begin{equation}
\label{FLRWmetric}
ds^2=-dt^2+a(t)^2 \left(\frac{dr^2}{1-k r^2}+r^2 d\Omega^2 \right)\, ,
\end{equation}
The field equations resulting from the action (\ref{theo}) are
\begin{equation}
\label{Friedmann1}
3F(H)=\kappa \rho+\Lambda \,,
\end{equation}
\begin{equation}
\label{Friedmann2}
-\frac{\dot{H}}{H}F'(H)=\kappa(\rho+P) \,,
\end{equation}
where
\begin{equation}
\label{FdeH}
F(H)= H^2+L^{-2}\sum_{n=3}^{\infty} (-1)^n\lambda_n \left(LH\right)^{2n}.
\end{equation}
and 
\begin{equation}
\label{dFdeH}
F'(H)\equiv \frac{dF(H)}{dH}\,,
\end{equation}
we have assumed $k=0$, $\kappa=8\pi G$, $H=\dot{a}/a$ is the Hubble parameter, $\rho$ represents the usual matter and radiation energy densities, $P$ is the pressure of the fluid, and $\Lambda$ is the cosmological constant. The terms $\lambda_n$ are dimensionless constants, and the value of $L$ represents the energy scale of the theory. Notice that when all the couplings $\lambda_n$ are null, we recover the standard Friedmann equations for General Relativity.

In order to build a specific model, we can separate (\ref{FdeH}) into its even-$n$ and odd-$n$ parts as follows
\begin{eqnarray}
\label{foddeven}
F(H) & = & H^2+\frac{\lambda}{L^{2}}\sum^{\infty}_{n=0}\lambda^{\text{even}}_n (-L^2H^2)^{2n+4} \nonumber  \\ 
     &   &  \,\,\,\,\,\,\,\, \,\, +\frac{\lambda}{L^{2}}\sum^{\infty}_{n=0}\lambda^{\text{odd}}_n (-L^2H^2)^{2n+3}.
\end{eqnarray}

Accordingly with \cite{Arciniega:2018tnn} and \cite{Arciniega:2020pcy}, we analize the model of $F(H)$ with $\lambda_n^{odd}=0$ and $\lambda_n^{even}=1/n!$. The specific form of the function $F(H)$ is
\begin{equation}
\label{model}
F(H)=H^2+\lambda H^8L^6e^{(HL)^4}
\end{equation}

In \cite{Arciniega:2020pcy} we have shown that the predictions for this model in the $r$ vs $n_s$ space are within the 2-$\sigma$ confidence levels reported by \citep{Planck2018}, as well as the constraints for the slow-roll parameters $\epsilon_1$, $\epsilon_2$, and $\epsilon_3$. In the light of this agreement with observational constraints, we have chosen our model as indicated in eq. (\ref{model}). Nevertheless, the behavior is representative of all the models constructed in this way.

Let us analyze the resulting cosmology thoroughly.

\section{History of the Universe with Geometric Inflation}
\label{secc.History}
The modified Friedmann equations (\ref{Friedmann1}) and (\ref{Friedmann2}) in Geometric Inflation for model (\ref{model}) are
\begin{equation}
\label{eq:Fried1}
3H^2(1+\lambda H^8L^6 e^{(HL)^4})=\kappa \rho+\Lambda , \\
\end{equation}
and
\begin{equation}
\label{eq:Fried2}
-\dot{H}\left\lbrace 2+4\lambda H^6L^6(2+H^4L^4) \right\rbrace e^{(HL)^4}=\kappa (\rho+P) .
\end{equation}
We can rewrite this last equation in terms of the acceleration as
\begin{equation}
\label{eq:acc}
\frac{\ddot{a}}{a}=H^2-\frac{\kappa(\rho +P)}{2+4\lambda H^6L^6e^{(2+H^4L^4)}}
\end{equation}
If $\lambda=0$, we recover the case of General Relativity, and the need for an inflationary potential with negative pressure arises. When $\lambda$ and $L$ are different from zero, we notice that the second term in (\ref{eq:acc}) could be diluted, obtaining positive values of the acceleration. This acceleration should be large enough to produce an exponential expansion at the beginning of the Universe and, at the same time, $H$ should not be so big that the inflationary period does violate the Planckian limits.

In order to explore this balance, we integrate the modified Friedmann equations. Given that the field equations are of second-order, and for positive values of $\lambda$, the initial value problem is well-possed.  We impose initial conditions in the same way as in General Relativity. We take the best fit values reported by \cite{Aghanim:2018eyx}, $\Omega_m^0=0.31$, $\Omega_r^0=8.4\times 10^{-5}$ and $\Omega_\Lambda=0.69$. In order to avoid interpreting this modification as some kind of dark fluid, the critical density can be redefined as
\begin{equation}
\label{eq:rhocrit}
\rho_{c}=\frac{3}{\kappa}F(H) .
\end{equation}
For the archetypical model that we are using, the critical density is $\rho_{c}=\frac{3}{\kappa}H^2(1+\lambda H^6L^6 e^{(HL)^4})$.  The usual $\rho_{c}$ can be recovered when $\lambda=0$, and by writing the critical density in this way, the energy scale at which the modification is relevant becomes explicit.

\subsection{Inflation}
As was demonstrated in \cite{LINDE1982389}, in order to build a viable inflationary period, it is necessary to obtain an almost constant evolution of the Hubble parameter during a large enough period of time to produce an exponential expansion able to safe three important puzzles, the flatness of the universe, the horizon problem and the absence of topological defects like magnetic monopoles. In order to avoid the problems presented in \cite{Guth:1980zm}, the evolution of $H$ should not be exactly constant so the exponential expansion can reach an end. 

In order to explore if the three problems can be solved by Geometric Inflation, we integrate the field equations (\ref{eq:Fried1}) and (\ref{eq:Fried2}). Initial conditions are set at the present day, $\Omega_m^0=0.31$, $\Omega_r^0=8.4\times 10^{-5}$ and $\Omega_\Lambda=0.69$  \cite{Aghanim:2018eyx}. This way, we have imposed the flatness condition and we will see what implications it has in our inflationary period. 

\begin{figure}
\begin{center}

\includegraphics[width=9.5cm]{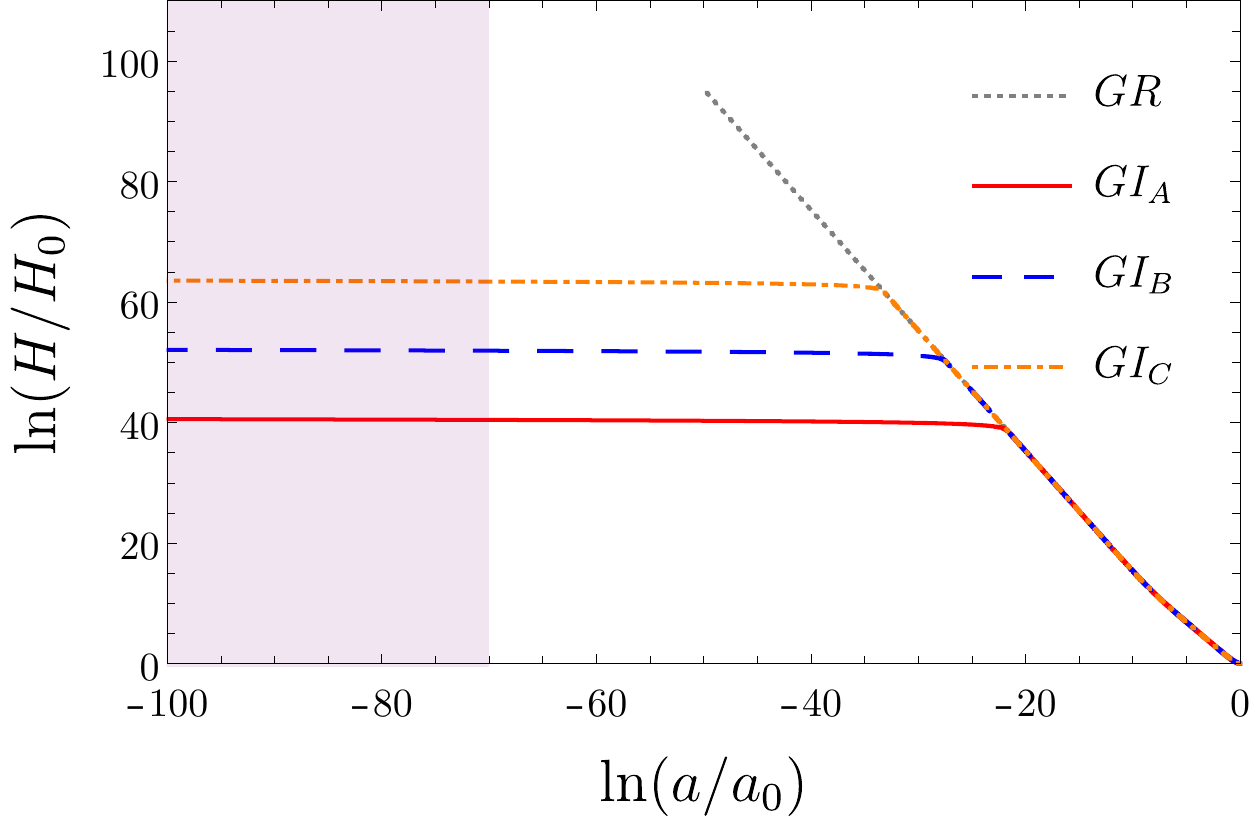}

\caption{Evolution of the Hubble parameter $H$ during the full history of the Universe. The grey dotted line represents the evolution in standard General Relativity. The red solid, blue dashed and orange dot-dashed lines show the evolution of $H$ in Geometric inflation for $L=1\times 10^{-17}H_0^{-1}$, $L=1\times 10^{-22}H_0^{-1}$, and $L=1\times 10^{-27}H_0^{-1}$, respectively. All cases have $\lambda=1$. The pink area represents the {\it forbiden zone} imposed by the Planck density. }
\label{Fig:Hfull}
\end{center}

\end{figure}

In Figure \ref{Fig:Hfull}, we plot the evolution of $H(a/a_0)$ for different values of the energy scale $L$. Labels $A$, $B$ and $C$ represent values $L=1\times 10^{-17}H_0^{-1}$, $L=1\times 10^{-22}H_0^{-1}$, and $L=1\times 10^{-27}H_0^{-1}$, respectively. We have fixed $\lambda=1$ for all the cases. We notice that the evolution of $H$ is quite similar to GR at late times. In early times, the evolution of $H$ becomes flat. The slow-roll condition, $-\dot{H}/H^2 \ll 1$, becomes a prediction of this theory. Notice that the contribution of radiation will be kept all along with the integration. In the same figure, the pink area represents the {\it forbidden zone} imposed by the Planck density.

\begin{figure}
\begin{center}
\includegraphics[width=9.5cm]{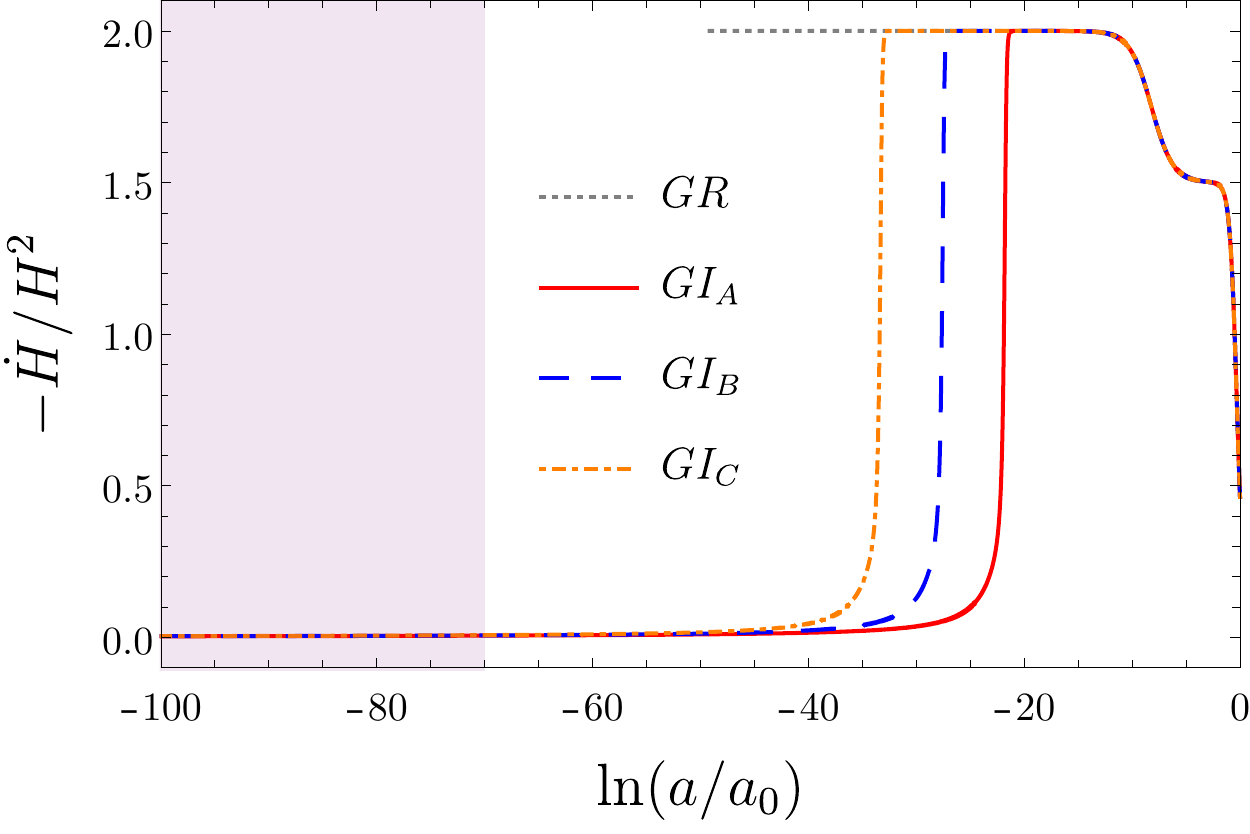}
\caption{Evolution of the slow-roll parameter $\epsilon_1=-\dot{H}/H^2$ during the full history of the Universe. The grey dotted line represents the evolution in standard General Relativity. The red solid, blue dashed and orange dot-dashed lines show the evolution of $H$ in Geometric inflation for $L=1\times 10^{-17}H_0^{-1}$, $L=1\times 10^{-22}H_0^{-1}$, and $L=1\times 10^{-27}H_0^{-1}$, respectively. All cases have $\lambda=1$. The pink area line represents the {\it forbiden zone} imposed by the Planck density.}
\label{Fig:Slow-roll}
\end{center}
\end{figure}

In order to show in a more precise way the evolution of the inflationary period and its graceful exit, in figure \ref{Fig:Slow-roll}, we show the evolution of the slow-roll parameter $\epsilon_1=-\dot{H}/H^2$ for the same values of the energy scale $L$. It can be clearly noticed, from figures \ref{Fig:Hfull} and \ref{Fig:Slow-roll}, that the duration of the exponential evolution, corresponding to the slow-roll behavior, can be sustained for even more than 60 $e$-folds, crossing the size of the universe when the Planck density is reached.  
Some concerns regarding this fact have been manifested in \cite{2020JCAP...12..040E}, where the authors argue that the frontier imposed by the Planck density would make it impossible to obtain enough number of $e-$folds during inflation to solve the three puzzles of inflation. This is an important aspect to solve.  We would like to argue about this, in the frame of Geometric inflation, and we will try to avoid assumptions that could rely on the standard interpretation of inflation in General Relativity.

The horizon problem force any kind of inflationary strategy to satisfy the geometric equation
\begin{equation}
e^{\mathcal{N}} > \frac{a_IH_I}{a_0H_0}
\end{equation}
where $\mathcal{N}$ represents the number of $e-$foldings,  $a_I$ and $H_I$ are the sizes of the universe, and the value of the Hubble parameter at the end of the inflationary epoch, $a_0$ and $H_0$ are the size of the universe and the Hubble value today, respectively. The most restrictive limit that we have for the exit of the exponential evolution is given by Big Bang Nucleosynthesis. The primordial abundances for light elements strongly depend on the expansion rate $H$, by that time the evolution of $H(Z)$ can not have strong differences with standard General Relativity. The number of $e-$folds should be $\mathcal{N} > 17$.  Model $A$ in figure  \ref{Fig:Hfull}, presents its graceful exit around the time of BBN. 

The hypothesis of the inflaton forces to end the exponential expansion to make the inflationary field decay into radiation, the era of radiation starts with that process.  Also, to reach the slow-roll behavior, that in terms of the Hubble parameter means that $\dot{H}\sim 0$, the conditions that the perturbations of the field have to reach implies an energy density around $\rho\sim 2\times 10^{16}GeV$, under the paradigm of the inflaton, what it is assumed in GR is that, by that time (or size of the universe) the inflaton was already decayed into radiation so the inflationary process should be ended. The predicted number of $e-$folds under such an scheme is $\mathcal{N} >62$.  

In the frame of GI, radiation is present all the time, at least, since the Planck density is reached.  The slow-roll condition, translated into $\dot{H}\sim 0$, is a prediction of the theory. The size of the universe at the exit of inflation is given by the value of the energy scale $L$. The thermal history of the universe, when it only depends on the size of the universe, could be reproduced given that the temperature is related to the size of the universe and not necessarily with the rate of the expansion.  The case $C$ in figures  \ref{Fig:Hfull} and \ref{Fig:Slow-roll},  represents the evolution of a universe with a graceful exit around the standard epoch for baryogenesis. The number of $e-$folds to fulfil the horizon condition is $\mathcal{N}\sim 27$. The exit of the inflationary period can be located at an energy scale lower than $10^9$ GeV, as suggested in \cite{Bedroya_2020}.

Regarding the flatness of the universe, as we have mentioned, such a condition was imposed in the present analysis. Nevertheless,  it is worth mentioning that, mathematically, the universe under this frame reaches the original singularity at an infinite time. If the value of the curvature were close to $\Omega_k \sim 1$, then the exponential expansion where $\dot{H}\sim 0$ would fulfill the flatness condition. The emergence of radiation will make the change to end the eternal exponential expansion and gracefully connect with the GR regime., there is no need to use an alternative mechanism to turn off the inflationary epoch. Finally, regarding the absence of topological defects like monopoles, the standard number of $e-$foldings required is $\mathcal{N}\sim 23$, this number can be easily provided by the three cases of GI explored here.

\subsection{Big Bang Nucleosynthesis}

Nucleosynthesis is perhaps the most restrictive test that every modification of gravity has to fulfill (see \cite{2010ARNPS..60..539P} and references therein). The geometrical measurement that is of our concern during this period is the expansion rate $H$.  Deviations in the evolution of the Hubble parameter could produce significant differences in the abundances of the light elements. A first test to demonstrate the viability of the theory is to show that the evolution of the Hubble parameter, $H$, is close to General Relativity during this epoch. 

For a Universe dominated by radiation, in GR, the expansion ratio $H$ and the temperature $T$ relate as $H^2=\kappa\Omega_r^0a^{-4} \propto T^4$. This relation should be corrected in Geometric inflation by $F(H)\propto T^4$. The freeze-out takes place when the weak interaction ratio, $\Gamma_{wk}$, is lower than the expansion ratio $H$, $\Gamma_{wk}<H$. For the archetypical model of GI in this paper, we can rewrite $F(H)=H^2(1+\lambda H^6L^6e^{(HL)^4})$, then the freeze-out will remain the same while the term ${1+\lambda H^6L^6e^{(HL)^4}}\sim 1$. This factor determines how close we are to the GR predictions for BBN and how significant the changes could be. 

\begin{figure}
\begin{center}
\includegraphics[width=9.5cm]{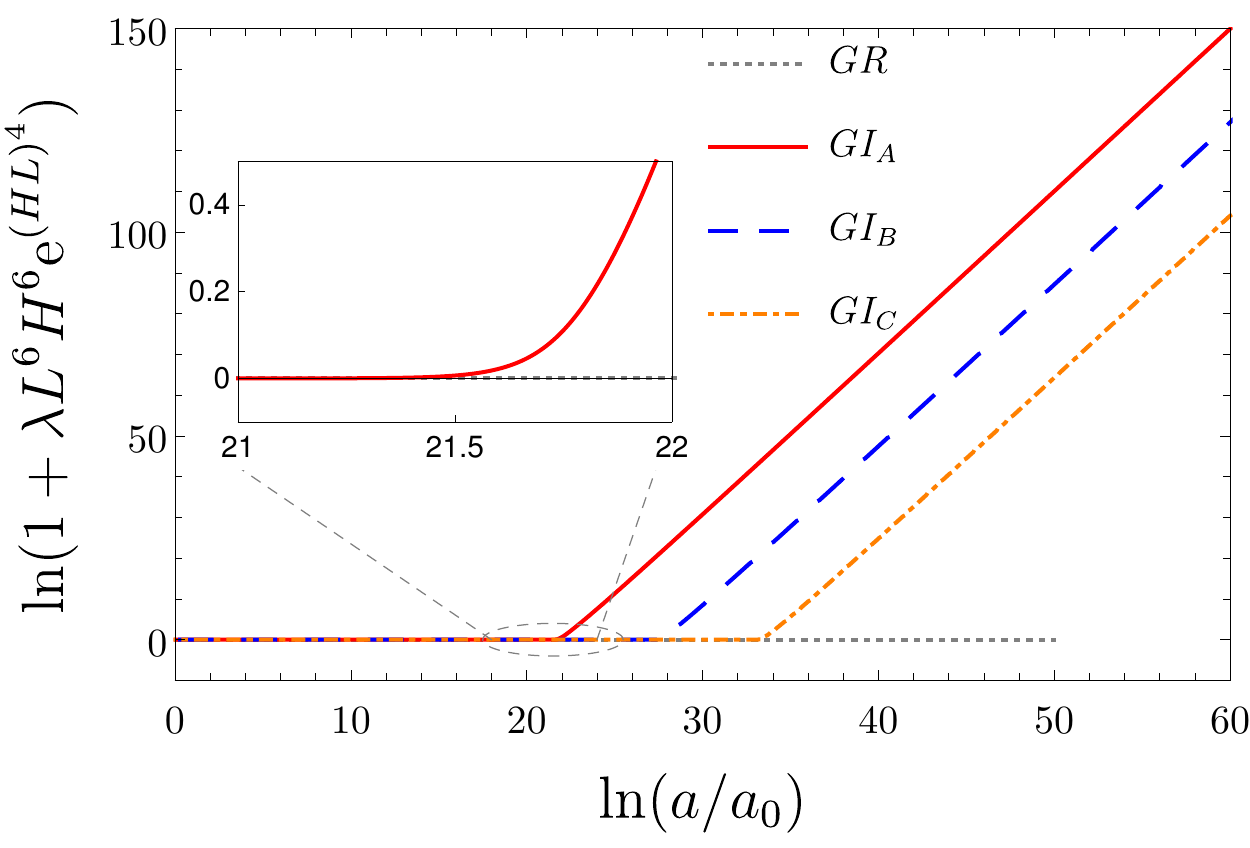}
\caption{Evolution of the Freeze-out factor. Big Bang Nucleosynthesis  is located around $z\approx 10^{10}$ or Ln$(a/a_0)\approx -23$. The grey dotted line represents the evolution in standard General Relativity. The red solid line, blue dashed lines and orange dot-dashed line show Geometric inflation for $L=1\times 10^{-17}H_0^{-1}$, $L=1\times 10^{-22}H_0^{-1}$ and $L=1\times 10^{-27}H_0^{-1}$ respectively. The inset in the plot shows the zoom for the case $A$.}
\label{Fig:Freeze}
\end{center}
\end{figure}
Figure \ref{Fig:Freeze} shows the evolution of this factor for the three values of $L$ that we are exploring; in this case, the $x$-axis represents ln$(a/a_0)$. The standard location of nucleosynthesis in GR is $z\sim 1\times 10^{10}$ or ln$(a/a_0)\approx -23$. As can be noticed from the figure, there is a clear transition of this factor to the one expected in GR. The freeze-out factor changes dramatically when we enter into the inflationary epoch. In figure \ref{Fig:Freeze} we show a close up of the evolution for one of the cases, as the value $1+\lambda H^6L^6e^{(HL)^4} \rightarrow 1$. The transition goes smoothly to the GR regime. With this simple test, we can conclude that, within the present framework, BBN constraints can be fulfilled at least as well as in GR. 

We have redefined the critical density as (\ref{eq:rhocrit}); this redefinition will have an impact on the densities of the matter and radiation components of the universe. In particular, the value of the baryonic density, $\Omega_b$, will be different in general. Depending on the values of the scale energy $L$ and $\lambda$, the value of $\Omega_b^{GI}$, will be different from the one in GR by a factor of $1+\lambda H^6L^6 e^{(HL)^4}$. The choice of $L$ and $\lambda$ could help to relax the missing baryons problem and still be consistent with the primordial abundances predicted by BBN. To have an idea about this change, if we approximate, the value of the Hubble parameter during BBN as $H/H_0=5\times10^{18}$, the energy scale $L=1\times 10^{-18}$ and the value of $\lambda=10$, then the value $\Omega_b^{GI}$ will be $10\%$ lower than the value predicted by GR. As we have shown, the value $L=1\times 10^{-18}H_0^{-1}$ could be within the range to pass the BBN constrictions. This way, the missing baryons problem could become a way to constrain both $L$ and $\lambda$. If according to the recent reports in the literature (see \cite{10.1093/mnras/staa3782} and references therein), the missing baryons have been already observed, all of them, still impose a constrain on the energy scale of the theory.

\subsection{Late-time acceleration}
\begin{figure}
\begin{center}
\includegraphics[width=9.5cm]{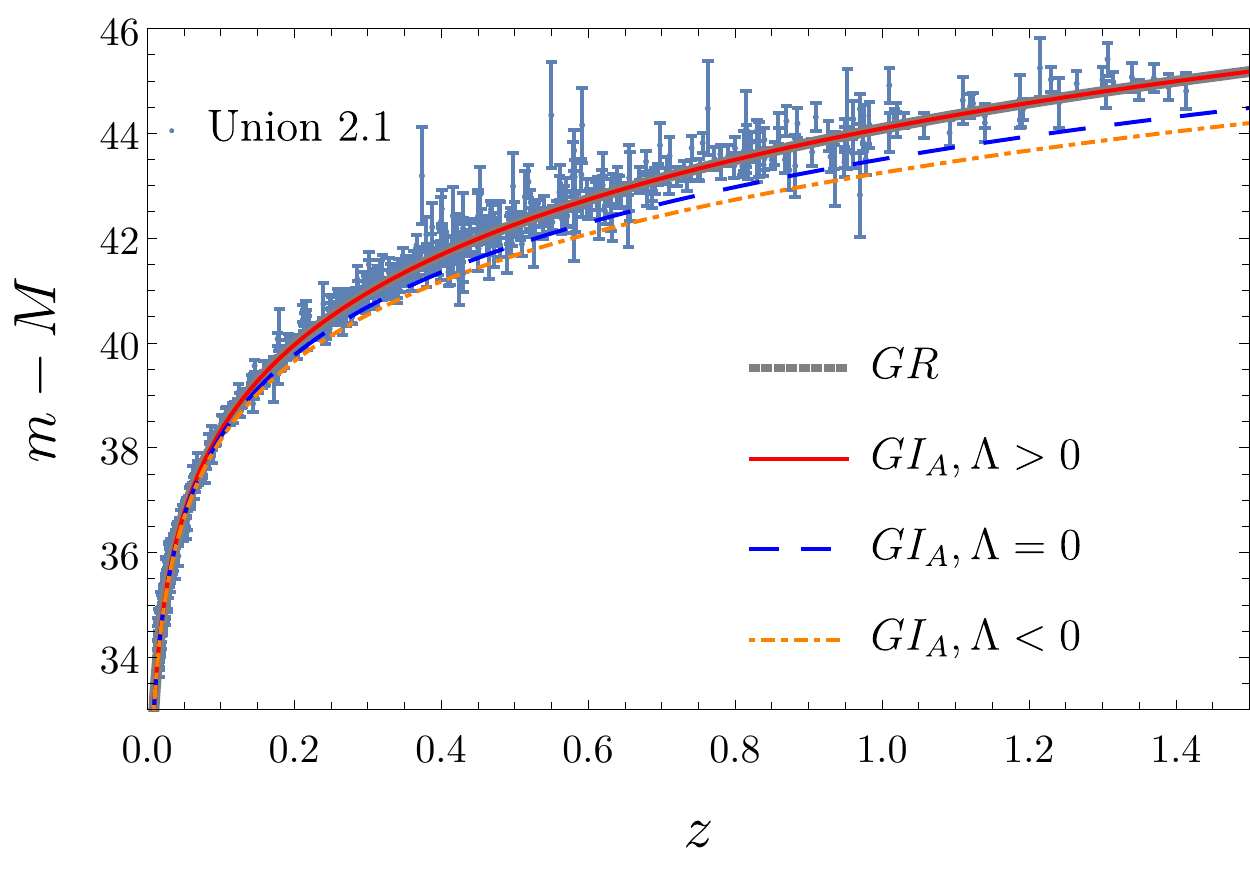}
\caption{$M-m$ for the UNION 2.1 Supernovae Ia collection. The solid grey line shows the standard evolution of GR with cosmological constant. Solid red lines depicts the evolution of GI case A with $L=1\times 10^{-17}H_0^{-1}$ and $\lambda=1$. Both cases take $(\Omega_m^0=0.3,\Omega_\Lambda^0=0.7)$. Blue dashed line and orange dot-dashed lines show the evolution for the same case (GIA) with $(\Omega_m^0=1,\Omega_\Lambda^0=0)$ and $(\Omega_m^0=0.5,\Omega_\Lambda^0=-1.5)$ respectively.}
\label{Fig:SNIa}
\end{center}
\end{figure}

With the evidence that we have shown, it is expected that the evolution at late time will be the same as in GR. Figure \ref{Fig:SNIa} depicts the evolution of the apparent magnitude $\mu$ of type Ia supernovae of the UNION 2.1 collection \cite{2010ApJ...716..712A}. We have included three cases, $\Omega_\Lambda>0$ with the best fit values of Planck 2018, $\Omega_\Lambda=0$ and $\Omega_\Lambda<0$. It can be noticed that the late-time evolution needs the cosmological constant to fit the data. We have performed some probes in the numerical runs to see if by changing the value of $\lambda$ and $L$ the data could be fitted taking $\Lambda=0$. We found no evidence to support such a hypothesis. 

There is another quantity that it is important to check to consider GI as a plausible cosmological model, the size of the sound horizon $r_s(z_*)$ during recombination. This quantity is given by
\begin{equation}
\label{eq:rs}
r_s(z)\equiv \int_{z_*}^\infty \frac{dz}{H(z)\sqrt{3R(z)+1}} ,
\end{equation}
where $z_*$ is the redshift to the last scattering surface, $z_*\sim 1089.95$ \cite{Aghanim:2018eyx} and $R(z)=\frac{3\Omega_r(z)}{4\Omega_b(z)}$. In GR, the upper limit of this integration is taken as $z\rightarrow \infty$ because the value of the integral (\ref{eq:rs}) becomes constant at high redshift. This limit has to change in the case of GI. Infinite redshift should be redefined by using different criteria: the end of inflation. The observational effect of this change will be imprinted in the angular size $\theta_*$ defined as
\begin{equation}
\theta_*=\frac{r_s(z_*)}{d_A(z_*)} ,
\end{equation}
where $d_A(z_*)$ is the comovil angular diameter distance at $z_*$. The distance $d_A(z)$, from the present day to $z_*$, does not change in GI. As we have shown, the evolution of $H(z)$ is the same as the one in GR for $z \leq z_{BBN}$. Table \ref{Table:Rs}, shows the predicted value of $r_s$ for GI compared to the expected value in GR. We have taken two different values of Ln$(a/a_0)$ as the new definition of $z_\infty$ considering the value of $\epsilon_1$. The standard criterium to consider that the Universe is still in its inflationary epoch is the value of $\epsilon_1\ll 1$. We consider two cases, if $\epsilon_1 = 0.10$ and $\epsilon_1 = 0.05$. We can notice the contribution to $r_s$ in GI when the evolution goes close to the inflationary epoch.

\begin{table}
\begin{center}
\begin{small}

\begin{tabular}{ |c|c|c|c|c|c|c|  }

 \hline
               & \multicolumn{2}{c|}{\,\,  Model  A \,\,  } & \multicolumn{2}{c|}{ \,\, Model  B \,\, }   &  \multicolumn{2}{c|}{ \,\, Model  C \,\, } \\
 \hline
 $\epsilon_1$  & 0.10 & 0.05 & 0.10 & 0.05 & 0.10 & 0.05 \\

 ln$(a/a_0)$   & -24.96 & -27.86 & -30.71 & -33.62 & -36.47 & -39.38 \\

 $\frac{r_s^{GI}(z_*)}{r_s^{GR}(z_*)}$ & 1.01 & 1.99 & 1.00 & 1.05 & 1.00 & 1.003\\

 \hline
\end{tabular}
\end{small}
 \caption{Value of $r_s(z_*)$ for Geometric Inflation in terms of the same quantity in General Relativity. We take two values of $\epsilon_1=-\dot{H}/H^2$, as the equivalent $z_\infty$. Cases $A$, $B$ and $C$ correspond to $L=1\times 10^{-17}H_0^{-1}$, $L=1\times 10^{-22}H_0^{-1}$ and $L=1\times 10^{-27}H_0^{-1}$ respectively. All the cases take $\lambda=1$.}
 \label{Table:Rs}
 \end{center}
\end{table}

\section{Conclusion}
We have presented the cosmological generic behavior of the new theory coined Geometric Inflation. On top of the Einsteinian characteristics demonstrated, we add with this work some features regarding its viability as a cosmological model. We have shown that the exponential evolution during the inflationary epoch can be reached without invoking an inflaton. Even more, we have shown that what is known as the "slow-roll" condition is a robust prediction within the frame of Geometric Inflation.  During the inflationary period, the exponential expansion is driven by a combination of geometry and radiation, the role of the inflaton to heat the Universe seems irrelevant within this frame. The exponential expansion, while $-\dot{H}/H^2\sim 0$, can provide enough number of $e$-folds to solve the horizon problem if the end of inflation is located at lower energy values than $10^9$ GeV. Given that the age of the universe in this case is infinite, the flatness problem can be solved. 

We have also shown that the restrictions on the evolution of the Hubble parameter needed to keep the primordial abundances can be easily fulfilled. It is important to mention that if the inflaton becomes unnecessary, the seeds for the structure formation could come from the metric perturbations of the theory. Such an analysis is beyond the scope of the present paper. Currently, the analysis reported in the literature \cite{Cisterna:2018tgx,Pookkillath_2020,Jim_nez_2021} take into account a partial contribution of the Lagrangian densities.  It is important to mention that higher order contributions into the lagrangian can produce instabilities at linear or beyond linear order. In case that at linear order such instabilities may be contained such problems could arise within perturbations at higher order.    

Finally, we have shown that the late-time acceleration is the same as in GR, and we provide some elements to constraint the energy scale of the theory, $L$, and the value of $\lambda$. It is important to mention that the energy scale $L$ and the value of $\lambda$ should not be considered as simple parameters, they are fundamental quantities of the theory.

With the evidence presented in the present work, we demonstrate that, at the background level, Geometric inflation plus the cosmological constant will fit late-time cosmological data as well as General Relativity, with the great advantage that Geometric Inflation provides an inflationary epoch at the early universe. These conclusions make us wonder about the role of this kind of modification in the accelerated epochs of the universe, one analysis regarding geometric late-time acceleration can be seen in \cite{Jaime-Arciniega_2021}.

\section*{Acknowledgments}
LGJ thanks Gustavo Arciniega for very fruitful discussions and suggestions during the development of the paper. The author acknowledges the financial support of SNI, CONACyT-140630 and PAPIIT IN120620.




\begin{thebibliography}{10}
  \expandafter\ifx\csname url\endcsname\relax
 \def\url#1{\texttt{#1}}\fi
\expandafter\ifx\csname urlprefix\endcsname\relax\def\urlprefix{URL }\fi
\expandafter\ifx\csname href\endcsname\relax
  \def\href#1#2{#2} \def\path#1{#1}\fi

\bibitem{Heisenberg_2019}
L.~Heisenberg, \href{http://dx.doi.org/10.1016/j.physrep.2018.11.006}{A
  systematic approach to generalisations of general relativity and their
  cosmological implications}, Physics Reports 796 (2019) 1–113.
\newblock \href {https://doi.org/10.1016/j.physrep.2018.11.006}
  {\path{doi:10.1016/j.physrep.2018.11.006}}.
\newline\urlprefix\url{http://dx.doi.org/10.1016/j.physrep.2018.11.006}

\bibitem{1980PhLB...91...99S}
A.~A. {Starobinsky}, {A new type of isotropic cosmological models without
  singularity}, Physics Letters B 91~(1) (1980) 99--102.
\newblock \href {https://doi.org/10.1016/0370-2693(80)90670-X}
  {\path{doi:10.1016/0370-2693(80)90670-X}}.

\bibitem{2007JETPL..86..157S}
A.~A. {Starobinsky}, {Disappearing cosmological constant in f( R) gravity},
  Soviet Journal of Experimental and Theoretical Physics Letters 86~(3) (2007)
  157--163.
\newblock \href {http://arxiv.org/abs/0706.2041} {\path{arXiv:0706.2041}},
  \href {https://doi.org/10.1134/S0021364007150027}
  {\path{doi:10.1134/S0021364007150027}}.

\bibitem{2007PhRvD..76f4004H}
W.~{Hu}, I.~{Sawicki}, {Models of f(R) cosmic acceleration that evade solar
  system tests}, Physical Review D 76~(6) (2007) 064004.
\newblock \href {http://arxiv.org/abs/0705.1158} {\path{arXiv:0705.1158}},
  \href {https://doi.org/10.1103/PhysRevD.76.064004}
  {\path{doi:10.1103/PhysRevD.76.064004}}.

\bibitem{Arciniega:2018fxj}
G.~Arciniega, J.~D. Edelstein, L.~G. Jaime, {Towards geometric inflation: the
  cubic case}, Phys. Lett. B 802 (2020) 135272.
\newblock \href {http://arxiv.org/abs/1810.08166} {\path{arXiv:1810.08166}},
  \href {https://doi.org/10.1016/j.physletb.2020.135272}
  {\path{doi:10.1016/j.physletb.2020.135272}}.

\bibitem{Arciniega:2018tnn}
G.~Arciniega, P.~Bueno, P.~A. Cano, J.~D. Edelstein, R.~A. Hennigar, L.~G.
  Jaime, {Geometric Inflation}, Phys. Lett. B 802 (2020) 135242.
\newblock \href {http://arxiv.org/abs/1812.11187} {\path{arXiv:1812.11187}},
  \href {https://doi.org/10.1016/j.physletb.2020.135242}
  {\path{doi:10.1016/j.physletb.2020.135242}}.

\bibitem{Arciniega:2019oxa}
G.~Arciniega, P.~Bueno, P.~A. Cano, J.~D. Edelstein, R.~A. Hennigar, L.~G.
  Jaime, {Cosmic inflation without inflaton}, Int. J. Mod. Phys. D 28~(14)
  (2019) 1944008.
\newblock \href {https://doi.org/10.1142/S0218271819440085}
  {\path{doi:10.1142/S0218271819440085}}.

\bibitem{Arciniega:2020pcy}
G.~Arciniega, L.~Jaime, G.~Piccinelli, {Inflationary predictions of Geometric
  Inflation}, Phys. Lett. B 809 (2020) 135731.
\newblock \href {http://arxiv.org/abs/2001.11094} {\path{arXiv:2001.11094}},
  \href {https://doi.org/10.1016/j.physletb.2020.135731}
  {\path{doi:10.1016/j.physletb.2020.135731}}.

\bibitem{2009arXiv0907.5424B}
D.~Baumann, {TASI Lectures on Inflation}, arXiv e-prints (2009)
  arXiv:0907.5424\href {http://arxiv.org/abs/0907.5424}
  {\path{arXiv:0907.5424}}.

\bibitem{Bueno2016b}
P.~Bueno, P.~A. Cano, V.~S. Min, M.~R. Visser, Aspects of general higher-order
  gravities, Phys. Rev. D 95, 044010 (2017) (Oct. 2016).
\newblock \href {http://arxiv.org/abs/1610.08519} {\path{arXiv:1610.08519}},
  \href {https://doi.org/10.1103/PhysRevD.95.044010}
  {\path{doi:10.1103/PhysRevD.95.044010}}.

\bibitem{Bueno2017}
P.~Bueno, P.~A. Cano, On black holes in higher-derivative gravities (Mar.
  2017).
\newblock \href {http://arxiv.org/abs/1703.04625} {\path{arXiv:1703.04625}},
  \href {https://doi.org/10.1088/1361-6382/aa8056}
  {\path{doi:10.1088/1361-6382/aa8056}}.

\bibitem{Planck2018}
Y.~Akrami, et~al., {Planck 2018 results. X. Constraints on inflation}, Astron.
  Astrophys. 641 (2020) A10.
\newblock \href {http://arxiv.org/abs/1807.06211} {\path{arXiv:1807.06211}},
  \href {https://doi.org/10.1051/0004-6361/201833887}
  {\path{doi:10.1051/0004-6361/201833887}}.

\bibitem{Aghanim:2018eyx}
N.~Aghanim, et~al., {Planck 2018 results. VI. Cosmological parameters}, Astron.
  Astrophys. 641 (2020) A6.
\newblock \href {http://arxiv.org/abs/1807.06209} {\path{arXiv:1807.06209}},
  \href {https://doi.org/10.1051/0004-6361/201833910}
  {\path{doi:10.1051/0004-6361/201833910}}.

\bibitem{LINDE1982389}
A.~Linde,
  \href{https://www.sciencedirect.com/science/article/pii/0370269382912199}{A
  new inflationary universe scenario: A possible solution of the horizon,
  flatness, homogeneity, isotropy and primordial monopole problems}, Physics
  Letters B 108~(6) (1982) 389--393.
\newblock \href {https://doi.org/https://doi.org/10.1016/0370-2693(82)91219-9}
  {\path{doi:https://doi.org/10.1016/0370-2693(82)91219-9}}.
\newline\urlprefix\url{https://www.sciencedirect.com/science/article/pii/0370269382912199}

\bibitem{Guth:1980zm}
A.~H. Guth, {The Inflationary Universe: A Possible Solution to the Horizon and
  Flatness Problems}, Phys. Rev. D 23 (1981) 347--356.
\newblock \href {https://doi.org/10.1103/PhysRevD.23.347}
  {\path{doi:10.1103/PhysRevD.23.347}}.

\bibitem{2020JCAP...12..040E}
J.~D. {Edelstein}, D.~{V{\'a}zquez Rodr{\'\i}guez}, A.~{Vilar L{\'o}pez},
  {Aspects of geometric inflation}, Journal of Cosmology and Astroparticle
  Physics 2020~(12) (2020) 040.
\newblock \href {http://arxiv.org/abs/2006.10007} {\path{arXiv:2006.10007}},
  \href {https://doi.org/10.1088/1475-7516/2020/12/040}
  {\path{doi:10.1088/1475-7516/2020/12/040}}.

\bibitem{Bedroya_2020}
A.~Bedroya, R.~Brandenberger, M.~Loverde, C.~Vafa,
  \href{http://dx.doi.org/10.1103/PhysRevD.101.103502}{Trans-planckian
  censorship and inflationary cosmology}, Physical Review D 101~(10) (May
  2020).
\newblock \href {https://doi.org/10.1103/physrevd.101.103502}
  {\path{doi:10.1103/physrevd.101.103502}}.
\newline\urlprefix\url{http://dx.doi.org/10.1103/PhysRevD.101.103502}

\bibitem{2010ARNPS..60..539P}
M.~{Pospelov}, J.~{Pradler}, {Big Bang Nucleosynthesis as a Probe of New
  Physics}, Annual Review of Nuclear and Particle Science 60 (2010) 539--568.
\newblock \href {http://arxiv.org/abs/1011.1054} {\path{arXiv:1011.1054}},
  \href {https://doi.org/10.1146/annurev.nucl.012809.104521}
  {\path{doi:10.1146/annurev.nucl.012809.104521}}.

\bibitem{10.1093/mnras/staa3782}
J.~Chaves-Montero, C.~Hernández-Monteagudo, R.~E. Angulo, J.~D. Emberson,
  \href{https://doi.org/10.1093/mnras/staa3782}{{Measuring the evolution of
  intergalactic gas from z = 0 to 5 using the kinematic Sunyaev–Zel’dovich
  effect}}, Monthly Notices of the Royal Astronomical Society 503~(2) (2020)
  1798--1814.
\newblock \href
  {http://arxiv.org/abs/https://academic.oup.com/mnras/article-pdf/503/2/1798/36678282/staa3782.pdf}
  {\path{arXiv:https://academic.oup.com/mnras/article-pdf/503/2/1798/36678282/staa3782.pdf}},
  \href {https://doi.org/10.1093/mnras/staa3782}
  {\path{doi:10.1093/mnras/staa3782}}.
\newline\urlprefix\url{https://doi.org/10.1093/mnras/staa3782}

\bibitem{2010ApJ...716..712A}
{Amanullah}, T.~{Supernova Cosmology Project}, {Spectra and Hubble Space
  Telescope Light Curves of Six Type Ia Supernovae at 0.511 < z < 1.12 and the
  Union2 Compilation}, Astrophysical Journal 716~(1) (2010) 712--738.
\newblock \href {http://arxiv.org/abs/1004.1711} {\path{arXiv:1004.1711}},
  \href {https://doi.org/10.1088/0004-637X/716/1/712}
  {\path{doi:10.1088/0004-637X/716/1/712}}.

\bibitem{Cisterna:2018tgx}
A.~Cisterna, N.~Grandi, J.~Oliva, {On four-dimensional Einsteinian gravity,
  quasitopological gravity, cosmology and black holes}, Phys. Lett. B 805
  (2020) 135435.
\newblock \href {http://arxiv.org/abs/1811.06523} {\path{arXiv:1811.06523}},
  \href {https://doi.org/10.1016/j.physletb.2020.135435}
  {\path{doi:10.1016/j.physletb.2020.135435}}.

\bibitem{Pookkillath_2020}
M.~C. Pookkillath, A.~D. Felice, A.~A. Starobinsky,
  \href{https://doi.org/10.1088/1475-7516/2020/07/041}{Anisotropic instability
  in a higher order gravity theory}, Journal of Cosmology and Astroparticle
  Physics 2020~(07) (2020) 041--041.
\newblock \href {https://doi.org/10.1088/1475-7516/2020/07/041}
  {\path{doi:10.1088/1475-7516/2020/07/041}}.

\bibitem{Jim_nez_2021}
J.~B. Jim{\'{e}}nez, A.~Jim{\'{e}}nez-Cano,
  \href{https://doi.org/10.1088/1475-7516/2021/01/069}{On the strong coupling
  of einsteinian cubic gravity and its generalisations}, Journal of Cosmology
  and Astroparticle Physics 2021~(01) (2021) 069--069.
\newblock \href {https://doi.org/10.1088/1475-7516/2021/01/069}
  {\path{doi:10.1088/1475-7516/2021/01/069}}.
\newline\urlprefix\url{https://doi.org/10.1088/1475-7516/2021/01/069}

\bibitem{Jaime-Arciniega_2021}
L.~G. Jaime, G.~Arciniega, A unified geometric description of the universe:
  from inflation to late-time acceleration without a cosmological constant, To
  be submited (July 2021).

\end{thebibliography}





\end{document}